# Free vibration and wave power reflection in Mindlin rectangular plates via exact wave propagation approach

Seyyedmostafa Mousavi Janbehsarayi [1], Arian Bahrami *,[2], Mansour Nikkhah Bahrami[1]

[1] Department of Mechanical Engineering, College of Engineering, University of Tehran, Tehran, Iran.
[2] Department of Mechanical Engineering, Eastern Mediterranean University, G. Magosa, TRNC Mersin 10, Turkey. E-mail: arian.bahrami@emu.edu.tr

**Abstract**

Reflection, propagation and energy analysis are crucially important in designing structures, especially plates. A thick plate is considered based on first order shear deformation theory. Wave Propagation Method (WPM) is employed to exactly derive resonant frequencies and wave power reflection from different classical boundary conditions. Firstly, the frequency results are compared with other literatures to validate the exact proposed wave solution in the present work. Then, wave analysis and benchmark results for natural frequencies are presented for six different combinations of boundary conditions. The results indicate that the wave power reflection of thick rectangular plates is quite complicated and an incident wave of a specific type gives rise to other types of waves except for simply supported boundary conditions where the reflected wave power does not depend on the system parameters.

# 1. Introduction

Rectangular plates have extensive application in engineering from Nanotechnology [1] to Aerospace [2] and Biomechanics [3] and many others. Their responses to an external excitation and energy transmission to their neighborhoods must be studied carefully to avoid any probable damage [4].

Many researches have contributed to study the vibration of thin plates such as Leissa's exact solution [5]. Yet, the dynamics of thick plates is quite complex due to the variation of shear deformation across the thickness and effect of inertia forces [6]. Precedent studies in this field done by Reissner [7] and Mindlin [8]. Mindline plate theory, also known as first-order shear deformation theory (FSDT), considers the distribution of shear deformation across the thickness as a linear function and solves the obtained three equations of the motion. Total deflection of the plate consists of bending deflection, shear contribution and angles of rotation. Based on which of the mentioned parameters are considered as fundamental variables, the strategy for deriving the equation(s) of the motion will be determined, considering the fact that reducing the fundamental variables, and consequently equations of the motion will simplify the solution [9,10]. Although numerous analytical and numerical methods have been presented, most of them have limited applications [11,12]. Higher order shear deformation theories (HSDT) involve higher-order expansion of the displacements. This assumption increases the number of unknowns; Murty's theory of HSDT [13] deals with 5, 7, 9 unknowns, Kant [14] with 6 unknowns, and Lo et al. [15] with 11 unknowns.

Numerical and semi-analytical methods come in handy when the complexity of a problem precludes the analytical approaches to be used. The FEM and Rayleigh-Ritz energy methods are two major procedures for solving the obtained equations of the motion. The proposed solutions based on FEM method are able to solve the vibration of moderately thick plates with any combinations of boundary conditions [16]. Yet, shear locking problem is one of their salient concerns due to coupling between bending and shear modes [17]. Recently, Senjanović et al. [18] proposed a shear-locking-free FEM method for vibration analysis of Mindlin plates using bending deflection as a potential function for the definition of total deflection and angles of cross-section rotations. The Rayleigh-Ritz energy method [19-21] and boundary characteristic orthogonal polynomials along with three-dimensional Ritz formulation [22,23] have been used for the free vibration analysis of thick plates with arbitrary boundary conditions. The accuracy of the results is sensitive to the assumed natural modes presented by set of orthogonal functions. Hashemi et al. [24] proposed an exact analytical Levy type solution for thick plates using FSDT. They investigated the free vibration of moderately thick rectangular plates for six combinations of boundary conditions. There are also several other methods for vibration analysis of Mindlin-Reissner plates such as cell-based smoothed radial point interpolation method [25], discrete singular convolution method [26].

In addition to the methods mentioned above, there exists an exact approach known as wave propagation method (WPM). WPM is a simple, non-iterative, and efficient method for obtaining the natural frequencies of a system. Instead of applying boundary conditions to equations of the motion, reflecting, transmitting and propagating waves will be investigated to determine the natural frequencies and mode shapes of a system. One of the advantages of this method is the ability to study the energy transmission to the neighbors. This is very important for designing structures, because the effect of vibration to the neighbors and bases at frequencies near the natural frequencies will be determined prior to the construction. Wave propagation approach has been utilized mainly for finding the natural frequencies of beams, thin plates, rectangular and circular shells, membranes, frames, Nano-materials, and composite structures. Study of transmission and reflection matrices in Euler-Bernoulli [27] and Timoshenko [28] beams are two cases of WPM method application in beam theories. Bahrami et al. [29] used modified wave approach to find the natural frequencies of non-uniform beams, using Euler-Bernoulli beam theory. In another work, Bahrami et al. [30] used WPM for free vibration of non-uniform rectangular membranes. Annular circular and sectorial membranes were studied in [31,32] using two dimensional wave propagation. Also, the nonlocal scale effect on buckling, vibration and wave reflection in beams has been studied in [33,34]. The authors showed that, in nanotubes, the reflected power of an incident wave, except for simply supported boundary condition, is dependent upon the small scale parameter and incident wave frequency. Moreover, Bahrami studied the free vibration, wave power transmission and reflection in multi-cracked nanobeams [35] and nanorods [36]. Furthermore, Bahrami and Teimourian [37] presented the small scale effect on vibration and wave power reflection in circular annular thin nanoplates. Recently, Ilkhani et al. [38] studied energy reflection and transmission in rectangular thin nanoplates. They showed that except for simply supported boundary condition, in other conditions, the obtained coefficients of the transmission matrix, and consequently the energy reflection is dependent on the non-dimensional frequency parameter of the incident wave, the non-dimensional nonlocal parameter, the thickness to length ratio and the number of half waves in length direction.

Reviewing the above acknowledged literature provides us the clue that there is no research conducted on wave analysis and investigation of the effect of thickness of the plate on wave motion, conversion and reflection in thick plates. In all previously done researches in wave analysis of structures, there were at most two waves [27-38] while here there are three waves, and this makes the problem more complicated to analyze. In the present paper, a new analytical approach to analyze the free vibration and wave reflection in thick rectangular plates is presented using wave propagation method. In section 2, the governing equations of the motion with free, simply supported, and clamped boundary conditions are developed; the equations of the motion are rewritten in a specific format compatible with Wave Propagation Method (WPM). In this section, first, the equations of the motion are used to derive the exact propagation matrix, then exact reflection matrices are derived for mentioned boundary conditions. In addition, the propagation and reflection matrices will be helpful for future works that has to do with wave power transmission and reflection in waveguide structures. In section 3, numerical results are presented and investigated. The results are compared with other literatures and exact benchmark results are presented for the natural frequency for various aspect ratios, thickness to length ratios, and boundary conditions. As they are considered to be exact results, other researchers can use them to verify their approximate solutions in future works. Finally, the behavior of the reflection matrices is discussed for different boundary conditions. These results are discussed thoroughly for different thickness to length ratios and frequency ranges. Various boundary conditions are also considered to analyze the wave power reflection at boundaries. These results depict the behavior of the reflection coefficients which shows the energy reflected and dissipated at boundaries.

## 2. Methodology

### 2.1 Governing equation of motion

The non-dimensional equations of motion based on Mindlin plate theory for a flat, isotropic, thick rectangular plate of length $a$, width $L$ and thickness $h$ as shown in Fig.1 are [24]:

$$\tilde{\psi}_{1,11} + \eta^2 \tilde{\psi}_{1,22} + \frac{v_2}{v_1}\left(\tilde{\psi}_{1,11} + \eta \tilde{\psi}_{2,12}\right) - \frac{12K^2}{\delta^2}\left(\tilde{\psi}_1 - \tilde{\psi}_{3,1}\right) = -\frac{\beta^2 \delta^2}{12v_1}\ddot{\tilde{\psi}}_1, \tag{1a}$$

$$\tilde{\psi}_{2,11} + \eta^2\tilde{\psi}_{2,22} + \frac{v_2}{v_1}\eta(\tilde{\psi}_{1,12} + \eta\tilde{\psi}_{2,22}) - \frac{12K^2}{\delta^2}(\tilde{\psi}_2 - \eta\tilde{\psi}_{3,2}) = -\frac{\beta^2\delta^2}{12v_1}\tilde{\psi}_2, \qquad (1b)$$

$$\tilde{\psi}_{3,11} + \eta^2\tilde{\psi}_{3,22} - (\tilde{\psi}_{1,1} + \eta\tilde{\psi}_{2,2}) = -\frac{\beta^2\delta^2}{12K^2 v_1}\tilde{\psi}_3 \qquad (1c)$$

where $\delta = \frac{h}{a}$ is the dimensionless thickness to length ratio, $\eta = \frac{a}{b}$ is aspect ratio, $\beta = \omega a^2\sqrt{\frac{\rho h}{D}}$ is the non-dimensional frequency parameter, and $K^2$ is the shear correction factor to acknowledge the fact that transverse shear strains are not independent of the thickness coordinate. Also, $v_1 = (1-v)/2$ and $v_2 = (1+v)/2$ where $v$ is Poisson's ratio, and $\psi_{k,ij}$ is $\frac{\partial^2 \psi_k}{\partial i \partial j}$. In the Eqs. (1) comma-subscript convention represents the partial derivatives with respect to the normalized coordinates $(X_1, X_2, X_3)$. $\tilde{\psi}_3$ is non-dimensional transverse displacement, $\tilde{\psi}_1$ and $\tilde{\psi}_2$ are non-dimensional slope due to bending alone in the respective planes which are defined by the following relations,

$$\tilde{\psi}_1(X_1, X_2) = \psi_1(x_1, x_2, t)e^{-i\omega t}, \qquad (2a)$$

$$\tilde{\psi}_2(X_1, X_2) = \psi_2(x_1, x_2, t)e^{-i\omega t}, \qquad (2b)$$

$$\tilde{\psi}_3(X_1, X_2) = \psi_3(x_1, x_2, t)e^{-i\omega t}/a \qquad (2c)$$

where $t$ is time, $\omega$ is the frequency parameter, $X_1$ and $X_2$ are dimensionless coordinates defined as $X_1 = \frac{x_1}{a}$, $X_2 = \frac{x_2}{b}$. The displacements along the $x_1$ and $x_2$ axes are denoted by $u$ and $v$ respectively, while displacement in the direction of non-deformed middle surface is denoted by $w$. In Mindlin plate theory[8], the displacement components are assumed to be given as:

$$u = -x_3\psi_1(x_1, x_2, t) \qquad (3a)$$

$$v = -x_3\psi_2(x_1, x_2, t) \qquad (3b)$$

$$w = \psi_3(x_1, x_2, t) \qquad (3c)$$

The resultant bending moments, twisting moments, and the transverse shear forces in dimensionless form are:

$$\widetilde{M}_{11} = -(\breve{\psi}_{1,1} + \nu\eta\breve{\psi}_{2,2})e^{i\omega t} \tag{4a}$$

$$\widetilde{M}_{22} = -(\eta\breve{\psi}_{2,2} + \nu\breve{\psi}_{1,1})e^{iwt} \tag{4b}$$

$$\widetilde{M}_{12} = \widetilde{M}_{21} = -\nu_1(\eta\breve{\psi}_{1,2} + \breve{\psi}_{2,1})e^{iwt} \tag{4c}$$

$$\widetilde{Q}_1 = -(\breve{\psi}_1 - \breve{\psi}_{3,1})e^{i\omega t} \tag{4d}$$

$$\widetilde{Q}_2 = -(\breve{\psi}_2 - \eta\breve{\psi}_{3,2})e^{i\omega t} \tag{4e}$$

where $\widetilde{M}_{11}$ and $\widetilde{M}_{22}$ are the bending moments, $\widetilde{M}_{12}$ is the twisting moment and $\widetilde{Q}_1$ and $\widetilde{Q}_2$ are shear forces, all per unit length. Three types of classical boundary conditions are presented classified for an edge parallel to the $X_2$-normalized axis in Table 1.

## 2.2 Wave propagation method

From wave point of view, free vibration of any subject can be considered as waves which are traveling along the body. At boundaries, some parts of these propagating waves are reflected in different types of waves, and the rest will be dissipated or transmitted to the neighbors. In order to use WPM method, we must write the governing equation of the motion and boundary conditions in matrix form. Consider the free vibration problem when no external load is applied to the plate. Three dimensionless functions $\breve{\psi}_1$, $\breve{\psi}_2$ and $\breve{\psi}_3$ may be represented in terms of three dimensionless potentials $\Psi_1$, $\Psi_2$ and $\Psi_3$ as follows [24]:

$$\breve{\psi}_1 = C_1\Psi_{1,1} + C_2\Psi_{2,1} - \eta\Psi_{3,2} \tag{5a}$$

$$\tilde{\psi}_2 = C_1\eta\Psi_{1,2} + C_2\eta\Psi_{2,2} + \Psi_{3,1} \tag{5b}$$

$$\tilde{\psi}_3 = \Psi_1 + \Psi_2 \tag{5c}$$

where parameters $C_1$ and $C_2$ are $C_1 = 1 - \frac{\alpha_2^2}{\nu_1\alpha_3^2}$, $C_2 = 1 - \frac{\alpha_1^2}{\nu_1\alpha_3^2}$. Also, $\alpha_1, \alpha_2$ and $\alpha_3$ can be obtained from Eqs. (5) as follows:

$$\alpha_1^2, \alpha_2^2 = \frac{\beta^2}{2}\left[\frac{\delta^2}{12}\left(\frac{1}{K^2\nu_1}+1\right) \mp \sqrt{\left(\frac{\delta^2}{12}\right)^2\left(\frac{1}{K^2\nu_1}-1\right)^2 + \frac{4}{\beta^2}}\right], \tag{6a}$$

$$\alpha_3^2 = \frac{12K^2}{\delta^2\beta^2}\alpha_1^2\alpha_2^2 = \frac{12K^2}{\delta^2}\left(\frac{\beta^2\delta^4}{144K^2\nu_1} - 1\right) \tag{6b}$$

By substituting Eqs. (5) into Eqs. (1), three coupled equations of motion will be uncoupled in terms of potential parameters as:

$$\Psi_{1,11} + \eta^2\Psi_{1,22} = -\alpha_1^2\Psi_1 \tag{7a}$$

$$\Psi_{2,11} + \eta^2\Psi_{2,22} = -\alpha_2^2\Psi_2 \tag{7b}$$

$$\Psi_{3,11} + \eta^2\Psi_{3,22} = -\alpha_3^2\Psi_3 \tag{7c}$$

Considering that two opposite-side edges are simply-supported at $X_1 = 0$ and $X_1 = 1$, the solution for Eqs. (7) can be obtained as:

$$\Psi_1 = [A_1\sin(\lambda_1 X_2) + A_2\cos(\lambda_1 X_2)]\sin(m\pi X_1) \tag{8a}$$

$$\Psi_2 = [A_3\sinh(\lambda_2 X_2) + A_4\cosh(\lambda_2 X_2)]\sin(m\pi X_1) \tag{8b}$$

$$\Psi_3 = [A_5\sinh(\lambda_3 X_2) + A_6\cosh(\lambda_3 X_2)]\cos(m\pi X_1) \tag{8c}$$

where $A_i$ are arbitrary constants. The dispersion relation for these set of solutions is calculated in terms of $\alpha_i$:

$$\alpha_1^2 = (m\pi)^2 + \eta^2\lambda_1^2, \qquad \alpha_2^2 = (m\pi)^2 - \eta^2\lambda_2^2, \qquad \alpha_3^2 = (m\pi)^2 - \eta^2\lambda_3^2 \qquad (9)$$

Substituting Eqs. (8) using Eqs. (5), and by doing some calculations, $\check{\psi}_1, \check{\psi}_2$ and $\check{\psi}_3$ can be redefined as:

$$\check{\psi}_1 = \left[A'_1 C_1 m\pi e^{i\lambda_1 X_2} + A'_2 C_1 m\pi e^{-i\lambda_1 X_2} + A'_3 C_2 m\pi e^{\lambda_2 X_2} + A'_4 C_2 m\pi e^{-\lambda_2 X_2} + A'_5 \eta\lambda_3 e^{\lambda_3 X_2} \right. \\ \left. + A'_6 \eta\lambda_3 e^{-\lambda_3 X_2}\right] \cos(m\pi X_1) \qquad (10a)$$

$$\check{\psi}_2 = \left[A''_1 C_1 \eta\lambda_1 e^{i\lambda_1 X_2} + A''_2 C_1 \eta\lambda_1 e^{-i\lambda_1 X_2} + A''_3 C_2 \eta\lambda_2 e^{\lambda_2 X_2} + A''_4 C_2 \eta\lambda_2 e^{-\lambda_2 X_2} + A''_5 m\pi e^{\lambda_3 X_2} \right. \\ \left. + A''_6 m\pi e^{-\lambda_3 X_2}\right] \sin(m\pi X_1) \qquad (10b)$$

$$\check{\psi}_3 = \left[A'''_1 e^{i\lambda_1 X_2} + A'''_2 e^{-i\lambda_1 X_2} + A'''_3 e^{\lambda_2 X_2} + A'''_4 e^{-\lambda_2 X_2}\right] \sin(m\pi X_1) \qquad (10c)$$

Undefined amplitude parameters in the first equation of Eq. (10a) are:

$$A'_1 = \frac{-iA_1 + A_2}{2}, A'_2 = \frac{iA_1 + A_2}{2}, A'_3 = \frac{A_3 + A_4}{2} \qquad (11\text{ a})$$

$$A'_4 = \frac{-A_3 + A_4}{2}, A'_5 = \frac{-A_5 - A_6}{2}, A'_6 = \frac{-A_5 + A_6}{2}$$

For the coefficients of other two equations in the Eqs. (10), the following relations are detected:

$$A''_1 = iA'_1, \quad A''_2 = -iA'_2, \quad A''_3 = A'_3 \qquad (11b)$$

$$A''_4 = -A'_4, \qquad A''_5 = A'_5, \qquad A''_6 = -A'_6$$

$$A'''_1 = A'_1, \qquad A'''_2 = A'_2, \qquad A'''_3 = A'_3, \qquad A'''_4 = A'_4$$

These relations help to rewrite all Eqs. (10) with the same coefficients for all of them. Using Eq. (11b), the rewritten form of Eqs. (10) would be:

$$\check{\psi}_1 = \left[A'_1 C_1 m\pi e^{i\lambda_1 X_2} + A'_2 C_1 m\pi e^{-i\lambda_1 X_2} + A'_3 C_2 m\pi e^{\lambda_2 X_2} + A'_4 C_2 m\pi e^{-\lambda_2 X_2} + A'_5 \lambda_3 \eta e^{\lambda_3 X_2} \right. \\ \left. + A'_6 \lambda_3 \eta e^{-\lambda_3 X_2}\right] \cos(m\pi X_1) \qquad (12a)$$

$$\tilde{\psi}_2 = [A'_1 iC_1\lambda_1\eta e^{i\lambda_1 X_2} - A'_2 iC_1\lambda_1\eta e^{-i\lambda_1 X_2} + A'_3 C_2\lambda_2\eta e^{\lambda_2 X_2} - A'_4 C_2\lambda_2\eta e^{-\lambda_2 X_2} + A'_5 m\pi e^{\lambda_3 X_2} \quad (12b)$$
$$- A'_6 m\pi e^{-\lambda_3 X_2}] \sin(m\pi X_1)$$

$$\tilde{\psi}_3 = [A'_1 e^{i\lambda_1 X_2} + A'_2 e^{-i\lambda_1 X_2} + A'_3 e^{\lambda_2 X_2} + A'_4 e^{-\lambda_2 X_2}] \sin(m\pi X_1) \quad (12c)$$

In Eqs. (12), there are six traveling waves detected in the plate, three positive- and three negative-going waves traveling in $X_2$ direction with different wave numbers $\lambda_i$. The positive- and negative-going waves are defined as bellow:

$$\boldsymbol{a}^+(x) = \begin{Bmatrix} A'_2 e^{-i\lambda_1 X_2} \\ A'_4 e^{-\lambda_2 X_2} \\ A'_6 e^{-\lambda_3 X_2} \end{Bmatrix}, \quad \boldsymbol{a}^-(x) = \begin{Bmatrix} A'_1 e^{i\lambda_1 X_2} \\ A'_3 e^{\lambda_2 X_2} \\ A'_5 e^{\lambda_3 X_2} \end{Bmatrix} \quad (13a,b)$$

In which $\boldsymbol{a}^+(x)$ and $\boldsymbol{a}^-(x)$ are positive and negative-going waves traveling along $X_2$ direction ,respectively as shown in Fig.1. The first waves in Eqs. (13a,b) are considered generally to be propagating waves, and the next two waves with $\lambda_2$ and $\lambda_3$ wavenumbers are attenuating waves.

### 2.3 Propagation and reflection matrices

Consider two points on a vibrating plate a distance $X_0$ apart in $X_2$ direction as shown in Fig. 2. Positive- and negative-going waves propagate from one point to another. Denoting them as Eqs. (13), they are related by:

$$\boldsymbol{a}^+(X + X^0) = \boldsymbol{f}^+(X)\boldsymbol{a}^+(X^0), \quad \boldsymbol{a}^-(X^0) = \boldsymbol{f}^-(X)\boldsymbol{a}^-(X + X^0) \quad (14a,b)$$

where $X^0 = (X_1^0, X_2^0, X_3^0)$ is an arbitrary point on the plate, $X = (X_1, X_2, X_3)$ is the position of any point relative to $X^0$ in $X_2$ direction, and $\boldsymbol{f}^+(X), \boldsymbol{f}^-(X)$ are propagation matrices in positive and negative directions, respectively. By using Eqs. (13) and (14), we can derive the propagation matrices as bellow:

$$f^+(X) = f^-(X) = \begin{bmatrix} e^{-i\lambda_1 X_2} & 0 & 0 \\ 0 & e^{-\lambda_2 X_2} & 0 \\ 0 & 0 & e^{-\lambda_3 X_2} \end{bmatrix} \tag{15}$$

From wave point of view, free vibration of any subject can be considered as waves which are traveling along the body. At boundaries, the incident wave $\boldsymbol{a}^+$ gives rise to the reflected wave $\boldsymbol{a}^-$, which are related by [27]

$$\boldsymbol{a}^- = \boldsymbol{r}\boldsymbol{a}^+ \tag{16}$$

The $\boldsymbol{a}^+$ and $\boldsymbol{a}^-$ are positive- and negative-going waves respectively, and **r** is reflection matrix. In order to find the reflection matrices. We will find the reflection matrix for simply supported, clamped, and free boundary conditions.

The equilibrium conditions for the simply supported boundary condition from Table 1 are:

$$\widetilde{M}_{22} = 0, \quad \check{\psi}_1 = 0, \quad \check{\psi}_3 = 0 \tag{17}$$

By using Eqs. (4), (12), (13), and (17) and doing some calculations, it results in:

$$[-C_1\lambda_1^2\eta^2 - C_1(m\pi)^2\nu]a_1^+ + [C_2\lambda_2^2\eta^2 - C_2(m\pi)^2\nu]a_2^+ + [m\pi\eta\lambda_3 - \lambda_3\eta m\pi\nu]a_3^+ +$$
$$[-C_1\lambda_1^2\eta^2 - C_1(m\pi)^2\nu]a_1^- + [C_2\lambda_2^2\eta^2 - C_2(m\pi)^2\nu]a_2^- + [m\pi\eta\lambda_3 - \eta\lambda_3 m\pi\nu]a_3^- = 0 \tag{18a}$$

$$C_1 m\pi a_1^+ + C_2 m\pi a_2^+ + \eta\lambda_3 a_3^+ + C_1 m\pi a_1^- + C_2 m\pi a_2^- + \eta\lambda_3 a_3^- = 0 \tag{18b}$$

$$a_1^+ + a_2^+ + a_1^- + a_2^- = 0 \tag{18c}$$

By rewriting Eqs. (18) in a matrix form as Eq. (16), one gets:

$$\boldsymbol{r}_S = -I \tag{19}$$

where $\boldsymbol{r}_s$ is the reflection matrix for simply supported boundary condition, "I" is $3 \times 3$ identity matrix.

In clamped boundary condition, the equilibrium conditions at the boundary are (Table 1):

$$\breve{\psi}_1 = 0, \quad \breve{\psi}_2 = 0, \quad \breve{\psi}_3 = 0 \tag{20}$$

Substituting Eqs. (12) into Eq. (20), with respect to Eqs. (13) yields

$$C_1 m\pi a_1^+ + C_2 m\pi a_2^+ + \eta \lambda_3 a_3^+ + C_1 m\pi a_1^- + C_2 m\pi a_2^- + \eta \lambda_3 a_3^- = 0 \tag{21a}$$

$$-iC_1\lambda_1\eta a_1^+ - C_2\lambda_2\eta a_2^+ - m\pi a_3^+ + iC_1\lambda_1\eta a_1^- + C_2\lambda_2\eta a_2^- + m\pi a_3^- = 0 \tag{21b}$$

$$a_1^+ + a_2^+ + a_1^- + a_2^- = 0 \tag{21c}$$

By rewriting Eqs. (21) in a matrix form as Eq. (16), one obtains:

$$\boldsymbol{r_c} = -\begin{bmatrix} C_1 m\pi & C_2 m\pi & \eta\lambda_3 \\ -iC_1\lambda_1\eta & -C_2\lambda_2\eta & -m\pi \\ 1 & 1 & 0 \end{bmatrix}^{-1} \begin{bmatrix} C_1 m\pi & C_2 m\pi & \eta\lambda_3 \\ iC_1\lambda_1\eta & C_2\lambda_2\eta & m\pi \\ 1 & 1 & 0 \end{bmatrix} \tag{22}$$

where $\boldsymbol{r_c}$ is the reflection matrix for the clamped boundary condition. After some mathematical calculations and simplifications, the components of the reflection matrix for clamped boundary condition can be determined as:

$$r_c(1,1) = \{c_2[(m\pi)^2 - \eta^2\lambda_2\lambda_3] + c_1[-(m\pi)^2 - i\eta^2\lambda_1\lambda_3]\}/\Sigma \tag{23a}$$

$$r_c(1,2) = -2c_2\eta^2\lambda_2\lambda_3/\Sigma \tag{23b}$$

$$r_c(1,3) = -2m\pi\eta\lambda_3/\Sigma \tag{23c}$$

$$r_c(2,1) = 2ic_1\eta^2\lambda_1\lambda_3/\Sigma \tag{23d}$$

$$r_c(2,2) = \{c_2[(m\pi)^2 + \eta^2\lambda_2\lambda_3] - c_1[(m\pi)^2 - i\eta^2\lambda_1\lambda_3]\}/\Sigma \tag{23e}$$

$$r_c(2,3) = 2m\pi\eta\lambda_3/\Sigma \tag{23f}$$

$$r_c(3,1) = 2im\pi c_1\eta\lambda_1(c_1 - c_2)/\Sigma \tag{23g}$$

$$r_c(3,2) = 2m\pi c_2\eta\lambda_2(c_1 - c_2)/\Sigma \tag{23h}$$

$$r_c(3,3) = \{-c_2[(m\pi)^2 + \eta^2\lambda_2\lambda_3] + c_1[(m\pi)^2 + i\eta^2\lambda_1\lambda_3]\}/\Sigma \tag{23j}$$

where $r_c(i,j)$ is the element in i-th row and j-th column, and $\Sigma$ is a parameter defined for the sake of simplicity and it is equal to:

$$\Sigma = c_2[-(m\pi)^2 + \eta^2 \lambda_2 \lambda_3] + c_1[(m\pi)^2 - i\eta^2 \lambda_1 \lambda_3] \tag{24}$$

Finally, for free boundary conditions, the equilibrium conditions from Table 1 are:

$$\widetilde{M}_{22} = 0, \quad \widetilde{M}_{21} = 0, \quad \widetilde{Q}_2 = 0, \tag{25}$$

Substituting Eqs.(12) into Eq. (25), and with respect to Eqs. (13), 13) yields:

$$[-C_1 \lambda_1^2 \eta^2 - C_1(m\pi)^2 v]a_1^+ + [C_2 \lambda_2^2 \eta^2 - C_2(m\pi)^2 v]a_2^+ + [m\pi\eta\lambda_3 - \lambda_3 \eta m\pi v]a_3^+$$
$$+ [-C_1 \lambda_1^2 \eta^2 - C_1(m\pi)^2 v]a_1^- + [C_2 \lambda_2^2 \eta^2 - C_2(m\pi)^2 v]a_2^- \tag{26a}$$
$$+ [m\pi\eta\lambda_3 - \eta\lambda_3 m\pi v]a_3^- = 0$$

$$-2iC_1 \lambda_1 \eta m\pi a_1^+ - 2C_2 \lambda_2 \eta m\pi a_2^+ + [-\lambda_3^2 \eta^2 - (m\pi)^2]a_3^+ + 2iC_1 \lambda_1 \eta m\pi a_1^- \tag{26b}$$
$$+ 2C_2 \lambda_2 \eta m\pi a_2^- + [\lambda_3^2 \eta^2 + (m\pi)^2]a_3^- = 0,$$

$$(-iC_1\lambda_1\eta + i\lambda_1\eta)a_1^+ + (-C_2\lambda_2\eta + \lambda_2\eta)a_2^+ - m\pi a_3^+ + (iC_1\lambda_1\eta - i\lambda_1\eta)a_1^- \tag{26c}$$
$$+ (C_2\lambda_2\eta - \lambda_2\eta)a_2^- + m\pi a_3^- = 0,$$

Eqs. (26) can be written in a matrix form as $\boldsymbol{a}^- = \boldsymbol{r}\boldsymbol{a}^+$:

$$\boldsymbol{r}_F = - \begin{bmatrix} -C_1\lambda_1^2\eta^2 - C_1(m\pi)^2 v & C_2\lambda_2^2\eta^2 - C_2(m\pi)^2 v & m\pi\eta\lambda_3 - \lambda_3\eta m\pi v \\ -2iC_1\lambda_1\eta m\pi & -2C_2\lambda_2\eta m\pi & -\lambda_3^2\eta^2 - (m\pi)^2 \\ -iC_1\lambda_1\eta + i\lambda_1\eta & -C_2\lambda_2\eta + \lambda_2\eta & -m\pi \end{bmatrix}^{-1} \times$$

$$\begin{bmatrix} -C_1\lambda_1^2\eta^2 - C_1(m\pi)^2 v & C_2\lambda_2^2\eta^2 - C_2(m\pi)^2 v & m\pi\eta\lambda_3 - \eta\lambda_3 m\pi v \\ 2iC_1\lambda_1\eta m\pi & 2C_2\lambda_2\eta m\pi & \lambda_3^2\eta^2 + (m\pi)^2 \\ iC_1\lambda_1\eta - i\lambda_1\eta & C_2\lambda_2\eta - \lambda_2\eta & m\pi \end{bmatrix} \tag{27}$$

where $\boldsymbol{r}_F$ is the reflection matrix for free boundary condition. In order to analyze the vibration systemically, we need to use Eqs, (14), (16) and, the obtained propagation and reflection matrices. In this way, we will be able to analyze the frequency response of the system by solving the following equation:

$$\begin{bmatrix} -\mathbf{I}_{3\times3} & \mathbf{r}_A & 0 & 0 \\ \mathbf{f}^+ & 0 & -\mathbf{I}_{3\times3} & 0 \\ 0 & -\mathbf{I}_{3\times3} & 0 & \mathbf{f}^- \\ 0 & 0 & \mathbf{r}_B & -\mathbf{I}_{3\times3} \end{bmatrix} \begin{bmatrix} \mathbf{a}^+ \\ \mathbf{a}^- \\ \mathbf{b}^+ \\ \mathbf{b}^- \end{bmatrix} = 0 \qquad (28)$$

where a and b are the positive- and negative-going waves at two opposite boundaries as shown in Fig.1. By setting the imaginary and real parts of the determinant of Eq. (28), which is the familiar characteristic equation for the Mindlin plate, for simply supported, clamped and free boundary conditions, the resonant frequencies of each of the mentioned boundary conditions will be obtained.

## 3. Numerical results and discussion

### 3.1 Validation of the present method

In order to investigate the accuracy and reliability of this study, present results are compared with literature. Table 2 presents the comparison study of the fundamental non-dimensional natural frequency of the present work with [5], [19], [23], [22], and [24] for different boundary conditions, thickness to length ratio, and aspect ratio of the plate. For defining different boundary conditions at edges symbolisms are considered, for example, SCSF indicates that the edges $x_1 = 0, x_2 = 0, x_1 = a,$ and $x_2 = b$ are simply-supported, clamped, simply-supported and free, respectively. For comparison, the following material and geometrical properties are considered: shear correction factor, $K^2 = 0.86667$; Young's modulus, $E = 1$ TPa; shear modulus, $G = E/[2(1 + v)]$; Poisson's ratio $v = 0.3$; the rectangular plate is considered to be square and non-square by using two different aspect ratios ($\eta = 1, 0.4$); thickness to length ratio is investigated for four different values ($\delta = 0.001, 0.01, 0.1, 0.2$). Values *n* and *m* show that the vibrating mode has *n* and *m* half-waves in the $x_1$ and $x_2$ directions, respectively; their values are considered to be ($m = 1, n = 1$) for the comparison of the first mode shape. Among the aforementioned references, [5] and [24] have used exact characteristic equation for thin and thick plates respectively; alternatively, [19], [23], and [22] have implemented approximate solutions. Leissa's well-known paper [5] proposed exact characteristic equation for thin plates, so in order to compare present results with available data in his paper, the value for thickness to length ration is considered to be $\delta = 0.001$ in $\eta = 1$ case and $\delta = 0.01$ in $\eta = 0.4$ case. Another way of comparing the present results with the results of thin plate is by taking into account the relevant assumptions for thin plates in which one of them would be letting $K^2 \to$ inf. The results for exact characteristic equations for thick plates available in [24] are compared with exact results of the present work and as it is seen that they show precise agreement. The results are also in good agreement with approximate solutions of [19], [23], and [22] where the two dimensional Ritz method and boundary characteristic orthogonal polynomials along with three dimensional Ritz method was implemented, respectively. The point for the latter comparison is that the shear correction factor employed in the present study, [24] and [22] is $K^2 = 0.86667$, in [19] is $K^2 = 5/6$, and in [23] is assumed to be $K^2 = \pi/12$. This affects the results slightly.

### 3.2 Frequency calculation

The non-dimensional natural frequencies can be calculated using the present method as illustrated in Fig. 3 for a plate with SCSC boundary conditions, thickness to length ratio $\delta = 0.15$, aspect ratio $\eta = 0.6$, shear correction factor $K^2 = 0.86667$, Poisson's ratio $v = 0.3$, and $m=1$. As shown in Fig. 3, in the wave propagation method, for evaluating the natural frequencies, the real and imaginary curves of determinant of Eq. (28) should meet zero simultaneously. It should be mentioned that in this figure the first point where imaginary and real parts meet zero simultaneously is not a natural frequency, yet it is a cut-off frequency where the type of waves changes to another one. The first point after cut-off frequency where both imaginary and real parts meet zero is the fundamental natural frequency in $x_2$- direction ($n$=1) and the second point is the second natural frequency in that direction ($n$=2) and so forth.

The Cut-off frequencies (COF) are frequencies where the type of wave changes for frequencies bellow and above them. In this study, two types of waves are detected; attenuating waves and propagating waves. There are three positive- and negative-going waves in $x_2$ direction defined by Eq. (13). It is obvious that if $\lambda_1$ is real, the first wave will be a propagating wave; otherwise, it is an attenuating wave. Besides, if $\lambda_2$ and $\lambda_3$ are real or imaginary, the second and the third type of waves will be attenuating or propagating waves, respectively. Fig. 4 depicts the dispersion relations given by Eq. (9) where the relation between wavenumbers and the frequency is presented. The material and geometrical properties are considered as aforementioned properties. In Fig. 4, three cut-off frequencies are detectable where at each COF, the type of one wave changes. It is noticeable that at low frequencies there are only attenuating waves and at high frequencies there are only propagating waves. At each COF, one of the waves converts to another type; all three times from an attenuating wave to a propagating wave, so after three COFs, all attenuating waves convert to propagating waves.

Table 3 presents the frequency benchmark results for different boundary conditions, aspect ratios, thickness to length ratios, and different numbers of half waves in each direction. The results are obtained for SCSC, SCSS, SSSS, SCSF, SFSS, SFSF boundary conditions. The aspect ratio is considered to have different values of $\eta = 0.6, 0.8, 1, 1.5, 2$, and the thickness to length ratio is considered to be $\delta = 0.01, 0.1, 0.2$. Also, the natural frequency is given for four different mode numbers (($m,n$)=(1,1), (1,2), (2,1), (2,2)). These results are regarded as the exact results for thick plate so they are an excellent database for future works.

From the results presented in Table 3, it can be observed that the lowest natural frequencies correspond to the plate subjected to less edge restraints. As the number of edge restraint increases, the natural frequency parameters also increase. Among all six combinations of the boundary conditions listed in Table 3, it can be seen that the lowest and highest values of the frequency parameters correspond to S–F–S–F and S–C–S–C cases, respectively. Moreover, increasing both values of the thickness to length ratio ($\delta$) and aspect ratio ($1/\eta$) result in a decrease in natural frequency in every mode shapes and boundary conditions.

### 3.3 Reflection analysis of waves in Mindlin plates

Waves which are incident upon boundaries may reflect in different types of waves. There are three waves of two different types traveling in a thick plate based on first order shear deformation theory. The power carried in a propagating wave is proportional to the square of the wave amplitude. Thus, the power reflected per unit incident power can be calculated by the square of the reflection coefficients $|r_{ij}|^2$[27]. So, it is possible to calculate the reflected power at boundaries by investigating the behavior of the reflection coefficients. Coefficients of reflection matrix can be viewed from a different perspective. The j-th column of the reflection matrices represents the reflection contribution of the j-th incident wave in 1st, 2nd,…, i-th reflected waves while the i-th row of the reflection matrices represents the reflection contribution of 1st, 2nd,..,j-th incident waves in the i-th reflected wave. In other words, $r_{ij}$ represents the reflection contribution of the j-th incident wave in the i-th reflected wave. The mechanical properties of the plate is considered as the previous section. It is shown that reflection coefficients generally depend on five parameters and they are independent of the aspect ratio of the plate, so the position of the boundaries is unimportant. The COFs has significant effects on the moduli of reflection coefficients as at these frequencies, the type of waves changes; they are distinctly visible in reflection coefficients graphs. The thickness to length ratio of the plate has substantial effects on the behavior of the reflection coefficients and this effect is investigated. There are two physical boundary conditions considered namely clamped, and simply-supported boundary conditions. Different boundary conditions show different responses to incident waves in which the reflected power varies from one boundary condition to another one, so for each of these two classical boundary conditions, the results are discussed and investigated separately. It is observed that a specific type of incident wave may induce reflected waves of other types, and these phenomena profoundly depend on the type of the boundary condition and frequency parameter.

### 3.3.1 Reflection coefficients for simply supported boundary conditions

As it can be seen by Eq. (19), the reflection coefficients for simply supported edge are constant and they are independent of five parameters $m$, $\delta$, $K$, $\beta$, and $\nu$. In other words, if a propagating wave alone or an attenuating wave alone is incident upon the simply supported boundary, no energy will dissipate and the reflected power of the same type of incident wave is the same as the incident wave power. Furthermore, if a propagating wave or an attenuating wave alone is incident upon the simply supported boundary, exclusively the same type of wave will be reflected and there will not be any other types of reflected waves. In other words, if a propagating wave alone is incident upon the simply supported boundary, there will not be any attenuating wave in the reflected wave. This is due to the fact that the reflection matrix has zero off-diagonal components.

### 3.3.2 Reflection coefficients for clamped boundary conditions

For simplification, we divided the frequency range into four regions; $0 < \beta < \beta_{c1}$, $\beta_{c1} < \beta < \beta_{c2}$, $\beta_{c2} < \beta < \beta_{c3}$, and $\beta > \beta_{c3}$. Subscripts show the cut-off frequencies, for example, $\beta_{c1}$ shows the first cut-off frequency. In clamped boundary condition, the reflection matrix coefficients depend on $m$, $\delta$, $K$, $\beta$, and $\nu$ as it is clear from Eqs. (23) and (24). In Fig. 5, the moduli of the reflection coefficients $|r_{ij}|$ are depicted in case of clamped boundary conditions for different values of $\delta$. ($\delta = 0.15, 0.175, 0.2, 0.225$).

Three cut-off frequencies are recognizable in this figure where at these frequencies the wave type of each entry of reflection matrix is transformed into a different one causing sharp jumps or drops. It is noticeable from Fig. 5 that the values of these three COFs are dependent on the thickness to length ratio parameter δ. As the thickness to length ratio parameter increases, these three COFs shift to the lower frequencies.

In the first frequency region $0 < \beta < \beta_{c1}$, $\lambda_1$ is pure imaginary while $\lambda_2$ and $\lambda_3$ are real suggesting the existence of three attenuating waves a₁, a₂, a₃. In this frequency range, the moduli of eight reflection coefficients decrease as the frequency parameter increases except for the modulus of reflection coefficient $r_{32}$ where there is a frequency below $\beta_{c1}$ in which the maximum modulus of reflection coefficient $r_{32}$ occurs. In the first frequency region, increasing the thickness to length ratio leads to reduction in six moduli of the reflection coefficients $r_{11}, r_{12}, r_{13}, r_{21}, r_{22}, r_{23}$ except for $r_{31}, r_{32}, r_{33}$ which increase by increasing the thickness to length ratio $\delta$. In other words, if attenuating waves with $\lambda_1$ $\lambda_2$, $\lambda_3$ wavenumbers are incident upon the clamped boundary in the first frequency region, the power of the attenuating reflected wave with $\lambda_3$ wavenumber increases with increasing the thickness to length ratio $\delta$.

In the second frequency region between $\beta_{c1}$ and $\beta_{c2}$, $\lambda_1$ changes its value from imaginary into real and the other two wavenumbers $\lambda_2, \lambda_3$ remain unchanged. This implies that in the second region $\beta_{c1} < \beta < \beta_{c2}$ there are two attenuating waves a2, a3 and one propagating wave a1. In this frequency region, the modulus of $r_{11}$ is independent of the frequency and its value is constant and unity. So, if a propagating wave with $\lambda_1$ wavenumber is incident upon a clamped boundary condition, the power of the propagating reflected wave with $\lambda_1$ wavenumber will be the same as the incident wave power. As shown in Fig. 5, above the first COF, as the frequency parameter increases, the modulus of the reflection coefficient $r_{12}$ increases then, there is a reduction till it reaches zero at the second COF $\beta_{c2}$. The moduli of reflection coefficients $r_{13}$ and $r_{23}$ decreases till it reaches zero at the second COF. The modulus of the reflection coefficient $r_{21}$ increases till it reaches a maximum value then, there is a reduction till it reaches zero at the second COF. The modulus of the reflection coefficient $r_{22}$ increases then, there is a reduction till it reaches one at the second COF $\beta_{c2}$. In other words, if an attenuating wave with $\lambda_2$ wavenumber is incident upon a clamped boundary at the second COF, the power of the attenuating reflected wave with $\lambda_2$ wavenumber is equal to the incident wave power. The modulus of the reflection coefficient $r_{31}$ increases from zero then, there is a reduction till it reaches the second COF $\beta_{c2}$. The modulus of the reflection coefficient $r_{32}$ increases then, there is a reduction till it reaches the second COF $\beta_{c2}$. The modulus of the reflection coefficient $r_{33}$ increases then, there is a reduction till it reaches one at the second COF $\beta_{c2}$. In other words, if an attenuating wave with $\lambda_3$ wavenumber is incident upon a clamped boundary at the second COF, the power of the attenuating reflected wave with $\lambda_3$ wavenumber is equal to the incident wave power. In this frequency range, by increasing the thickness to length ratio, the moduli of all reflection coefficients increase except $r_{13}, r_{23}, r_{11}$ and $r_{21}$ where there is a reduction in three reflection coefficients $r_{13}, r_{23}, r_{21}$ and the modulus of reflection coefficient $r_{11}$ is independent of the thickness to length ratio.

In the third frequency region $\beta_{c2} < \beta < \beta_{c3}$, $\lambda_3$ becomes pure imaginary and the other two wavenumbers $\lambda_1, \lambda_2$ remain unchanged indicating two propagating waves a₁, a₃ and one attenuating wave a₂. In this frequency region, immediately after $\beta_{c2}$, as the frequency parameter increases, the moduli of the reflection coefficients $r_{11}$, $r_{32}$, $r_{33}$ decrease while the moduli of the reflection coefficients $r_{13}$, $r_{21}$, $r_{23}$ increase till $\beta_{c3}$. The situation is different for the reflection coefficients $r_{12}$, $r_{31}, r_{22}$. The modulus of reflection coefficient $r_{12}$ increases till it reaches a maximum value then, there is a reduction till it reaches zero at the third COF. The modulus of reflection coefficient $r_{31}$ decreases till it reaches a minimum value then, there is an increase till the third COF. The modulus of reflection coefficient $r_{22}$ is always unity in the third frequency region. In this frequency range, by increasing the thickness to length ratio, the moduli of all reflection coefficients decrease except $r_{13}, r_{23}, r_{22}$ and $r_{21}$ where there is an increase in three reflection coefficients $r_{13}, r_{23}$ and $r_{21}$ and the modulus of reflection coefficient $r_{22}$ is independent of the thickness to length ratio.

In the fourth frequency region $\beta > \beta_{c3}$, above the third COF, $\lambda_2$ becomes pure imaginary and the other two wavenumbers $\lambda_1, \lambda_3$ remain unchanged indicating three propagating waves a₁, a₂, a₃. In this frequency region, as the frequency parameter increases in the fourth region, the moduli of the reflection coefficients $r_{13}$, $r_{21}$, $r_{23}$, $r_{31}$, $r_{32}$ decrease and asymptote to the value of zero. In other words, the propagating reflected wave power of a propagating incident wave approaches zero in high frequency region. Also, as the frequency parameter increases in the fourth region, the moduli of the reflection coefficients $r_{11}$, $r_{22}$, $r_{33}$ increase and asymptote to the value of one. In other words, the propagating reflected wave power of a propagating incident wave is equal to the incident wave power in high frequency region. Moreover, as the frequency parameter increases in the fourth region, the modulus of the reflection coefficient $r_{12}$ increase and asymptote to the value of two. In other words, if a propagating wave with $\lambda_2$ wavenumber $a_2^+$ is incident upon a clamped boundary, the power of the reflected wave with $\lambda_1$ wavenumber $a_1^-$ becomes four times its incident wave power in high frequency region. In this frequency range, by increasing the thickness to length ratio, the moduli of all reflection coefficients increase except $r_{21}, r_{31}, r_{32}, r_{13}, r_{23}$ where there is a reduction in these five reflection coefficients.

## 4. Conclusion

The wave propagation method is implemented to analyze the energy reflection, propagation, and free vibration of thick plates based on Mindlin plate theory. First of all, the accuracy of the presented method is validated by comparing the obtained resonant frequencies with available literatures. Benchmark results for natural frequencies are presented for various thickness to length ratios, aspect ratios, numbers of half waves, and various combinations of boundary conditions. In future works, these results can be an excellent database to verify approximate or other analytical solutions as they are regarded as exact solutions. The main discussion of the present work is allocated to the investigation of wave power reflection, dissipation and conversion for two physical boundary conditions (clamped, and simply-supported). The $3 \times 3$ reflection matrices are obtained for each boundary conditions creating nine reflection elements $(r_{11}, r_{r_{12}}, \ldots r_{32}, r_{33})$, four frequency ranges based on three COFs, and three incident waves with $\lambda_1, \lambda_2,$ and $\lambda_3$ wavenumbers. For each region, the behavior of these nine arrays in the reflection matrix are determined. The three waves were either attenuating (at low frequencies) or propagating waves (at high frequencies). After each COF, one of the waves is converted to another type and every time from an attenuating wave to a propagating wave, so after three COFs all three waves change into the propagating waves. This step by step conversion has substantial impact on the response of the plate and energy conduction in the plate. It is shown that how each wave's energy contributes in another type of wave in the process of reflection and one type of wave induces another type of wave. The results show that the reflection from simply supported boundary condition does not depend on the system parameters, but it is dependent on the system parameters in case of clamped boundary conditions.

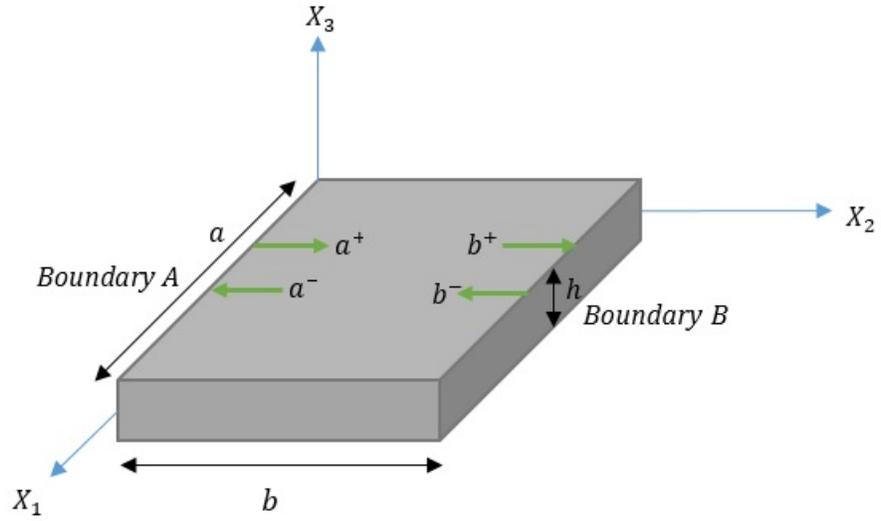

Figure. 1: A Mindlin plate with coordinate convention and waves.

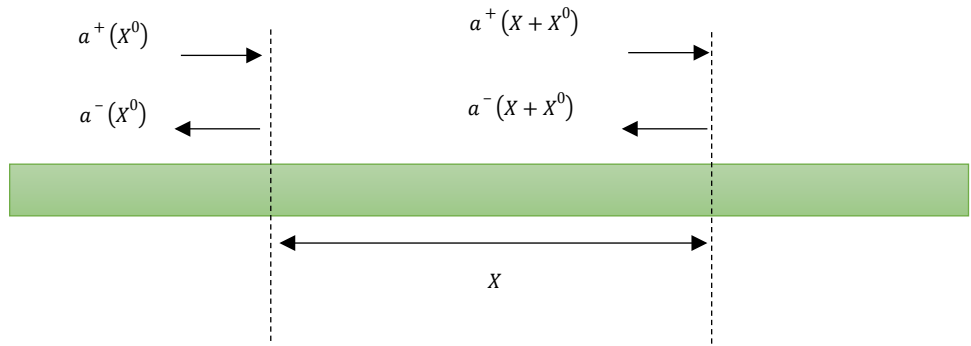

Figure. 2: A lateral view of Mindlin plate representing positive- and negative-going propagating waves.

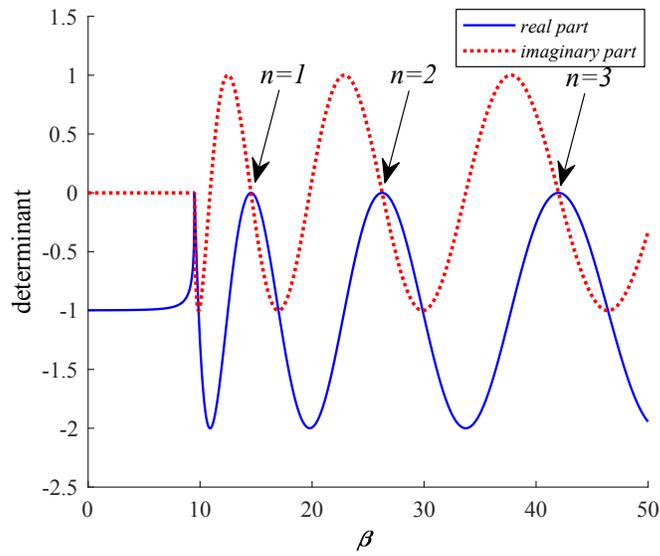

Figure. 3: Real and imaginary parts of determinant of Eq. (28)

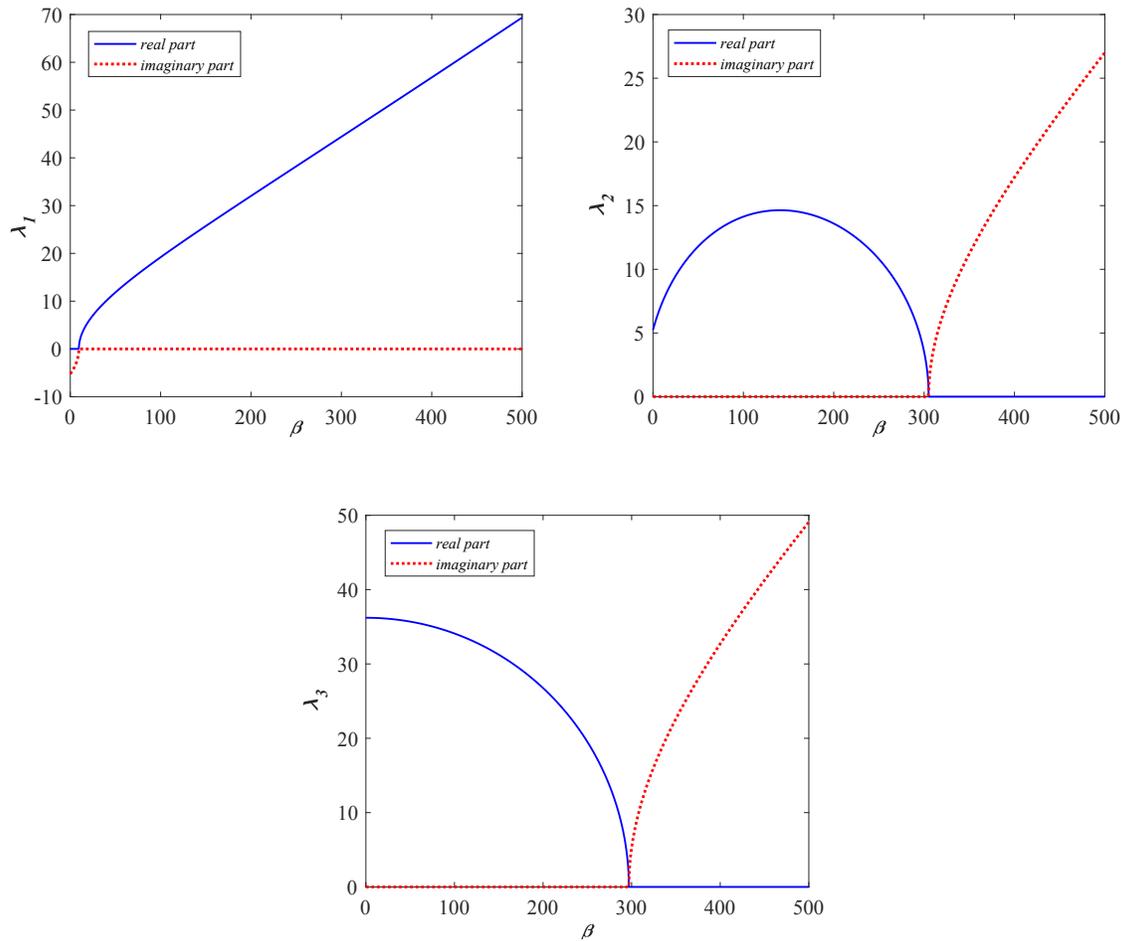

Figure. 4: Wavenumbers against frequency parameter using dispersion relation showing cut off frequencies

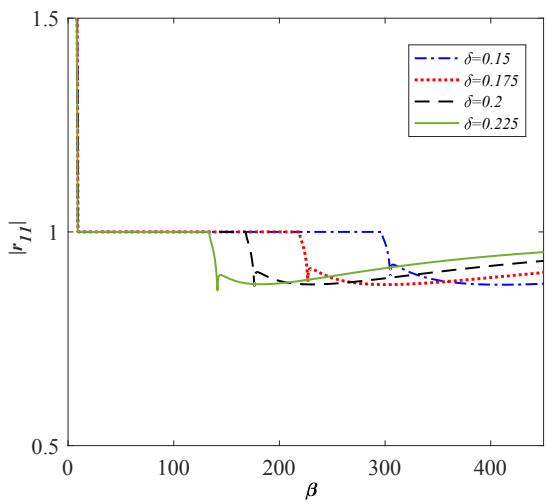
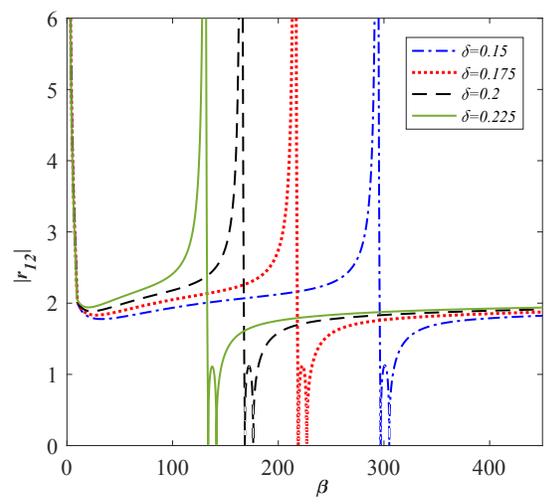
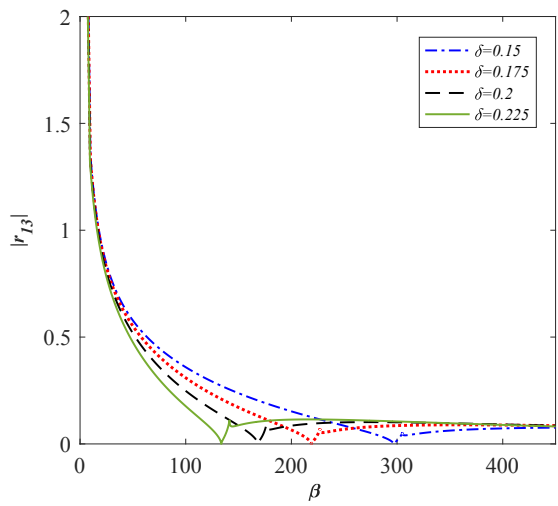
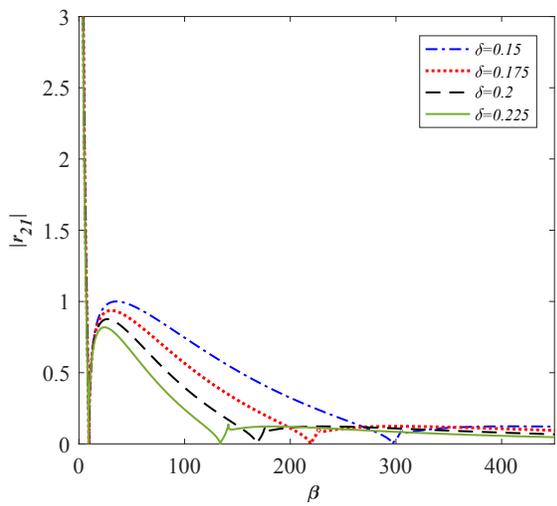

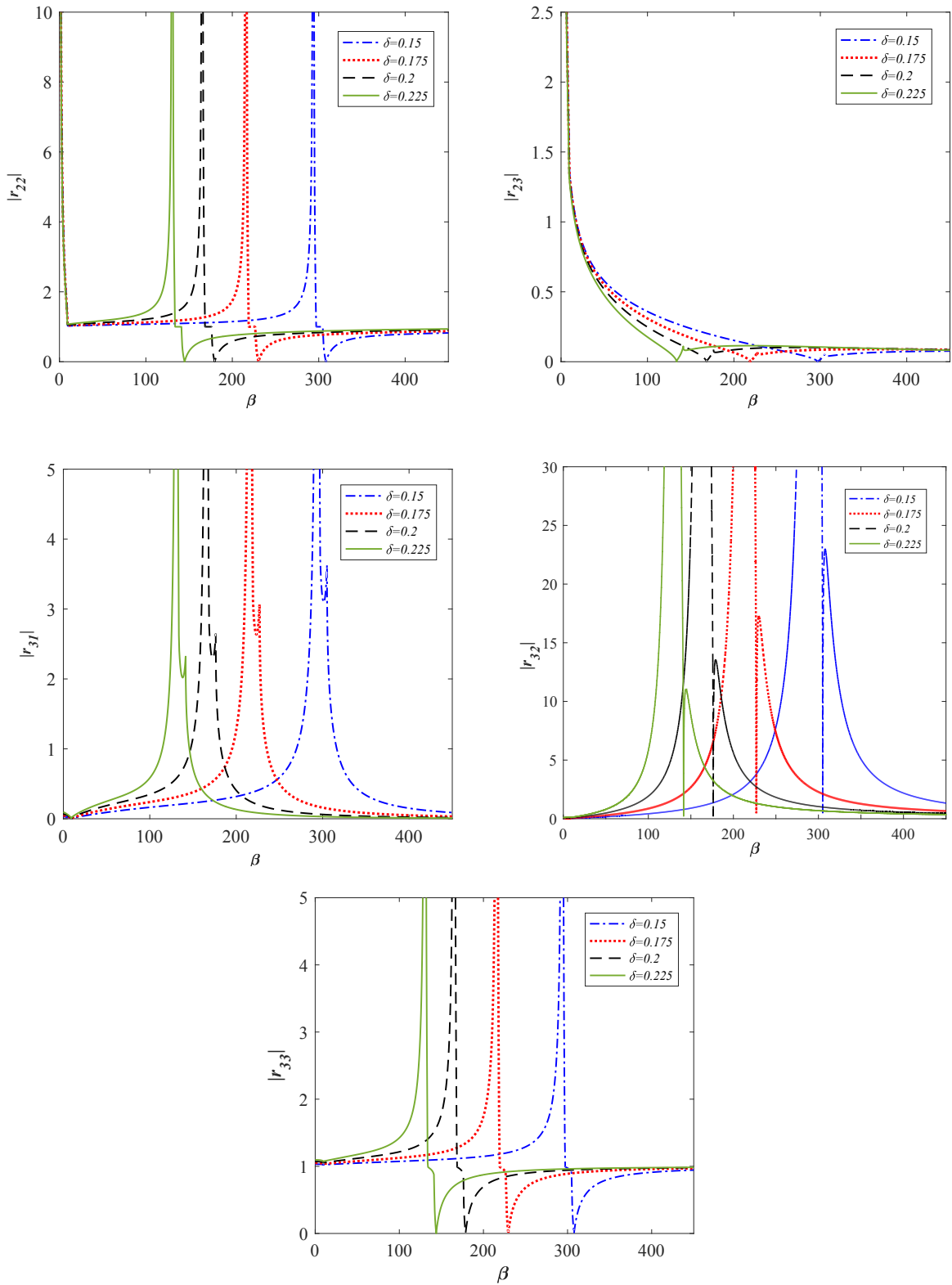

Figure. 5: The moduli of nine reflection coefficients against frequency parameter for different thickness to length ratios in case of clamped boundary condition

Table 1. Equilibrium conditions for different boundary conditions of Mindlin plates

| Boundary conditions | Equilibrium conditions |
|---|---|
| simply-supported edge | $\widetilde{M}_{22} = \tilde{\psi}_1 = \tilde{\psi}_3 = 0$ |
| clamped edge | $\check{\psi}_1 = \check{\psi}_2 = \check{\psi}_3 = 0$ |
| free edge | $\widetilde{M}_{22} = \widetilde{M}_{21} = \widetilde{Q}_2 = 0$ |

Table2 . Comparison of the fundamental frequencies $\beta = \omega a^2 \sqrt{\frac{\rho h}{D}}$ for Mindlin plates with different boundary conditions, aspect ratios, and thickness to length ratios

| | | Reference | SCSC | SCSS | SSSS | SCSF | SSSF | SFSF |
|---|---|---|---|---|---|---|---|---|
| $\eta = 1$ | $\delta = 0.001$ | Present[a] | 28.9505 | 23.6463 | 19.7391 | 12.6862 | 11.6838 | 9.6313 |
| | | Ref[24][a] | 28.9505 | 23.6463 | 19.7391 | 12.6863 | 11.6837 | 9.6311 |
| | | Ref[5] | 28.9509 | 23.6463 | 19.7392 | 12.6874 | 11.6845 | 9.6314 |
| | | Ref[19][b] | 28.9515 | 23.6456 | 19.7392 | 12.6854 | 11.6925 | 9.6406 |
| | | Ref[23][c] | 28.9475 | 23.6354 | 19.7392 | 12.6862 | 11.6978 | 9.6327 |
| | | Ref[22][c] | 28.9465 | 23.6325 | 19.7392 | 12.6868 | 11.6981 | 9.6354 |
| | $\delta = 0.1$ | Present[a] | 26.7369 | 22.4260 | 19.0840 | 12.2606 | 11.3808 | 9.4458 |
| | | Ref[24][a] | 26.7369 | 22.4260 | 19.0840 | 12.2606 | 11.3810 | 9.4458 |
| | | Ref[19][b] | 26.6687 | 22.4136 | 19.0651 | 12.2492 | 11.3727 | 9.4403 |
| | | Ref[23][c] | 26.6676 | 22.4336 | 19.898 | 12.2546 | 11.3789 | 9.4354 |
| | | Ref[22][c] | 26.6693 | 22.4365 | 19.0582 | 12.2573 | 11.3797 | 9.4462 |
| | $\delta = 0.2$ | Present[a] | 22.5099 | 19.7988 | 17.5055 | 11.3931 | 10.7218 | 8.9997 |
| | | Ref[24][a] | 22.5099 | 19.7988 | 17.5055 | 11.3931 | 10.7218 | 9.9998 |
| | | Ref[19][b] | 22.3596 | 19.7037 | 17.4485 | 11.3619 | 106987 | 8.9833 |
| | | Ref[23][c] | 22.4687 | 19.7348 | 17.5264 | 11.3765 | 10.7056 | 9.0010 |
| | | Ref[22][c] | 22.4569 | 19.7389 | 17.5351 | 11.3731 | 10.7102 | 9.0126 |
| $\eta = 0.4$ | $\delta = 0.01$ | Present[a] | 12.1316 | 11.7476 | 11.4464 | 10.1852 | 10.1222 | 9.7572 |
| | | Ref[24][a] | 12.1316 | 11.7476 | 11.4464 | 1018.48 | 10.1222 | 9.7569 |
| | | Ref[5] | 12.1347 | 11.7502 | 11.4487 | 10.1888 | 10.1259 | 9.7600 |
| | | Ref[19][b] | 12.1334 | 11.7554 | 11.4501 | 10.1854 | 10.1235 | 9.7583 |
| | | Ref[22][c] | 12.1328 | 11.7556 | 11.4508 | 10.1862 | 10.1238 | 9.7591 |
| | $\delta = 0.1$ | Present[a] | 11.8438 | 11.4978 | 11.2261 | 9.9876 | 9.9311 | 9.5816 |
| | | Ref[24][a] | 11.8438 | 11.4978 | 11.2260 | 9.9871 | 9.9310 | 9.5814 |
| | | Ref[19][b] | 11.8465 | 11.5026 | 11.2265 | 9.0008 | 9.9323 | 9.5832 |
| | | Ref[22][c] | 11.8456 | 11.5048 | 11.2235 | 9.0012 | 9.9324 | 9.5838 |
| | $\delta = 0.2$ | Present[a] | 11.1138 | 10.8485 | 10.6308 | 9.4910 | 9.4472 | 9.1314 |
| | | Ref[24][a] | 11.1138 | 10.8485 | 10.6307 | 9.9410 | 9.4470 | 9.1313 |
| | | Ref[19][b] | 11.1142 | 10.8498 | 10.6318 | 9.9425 | 9.44.85 | 9.1324 |

[a] Shear correction factor $K^2 = 0.86667$
[b] Shear correction factor $K^2 = 5/6$
[c] Shear correction factor $K^2 = \pi^2/12$

Table3. Natural frequencies, $\beta = \omega a^2 \sqrt{\frac{\rho h}{D}}$ for rectangular Mindlin plate

| δ | 1/η | SCSC | SCSS | SSSS | SCSF | SFSS | SFSF |
|---|---|---|---|---|---|---|---|
| | | (m=1,n=1) | | | | | |
| 0.01 | 0.6 | 67.9041 | 50.5219 | 37.2603 | 18.5719 | 14.4584 | 9.5357 |
| | 0.8 | 41.1303 | 32.0583 | 25.2794 | 14.4966 | 12.6204 | 9.5861 |
| | 1 | 28.9250 | 23.6327 | 19.7322 | 12.6728 | 11.6746 | 9.6270 |
| | 1.5 | 17.3650 | 15.5730 | 14.2525 | 10.9682 | 10.6654 | 9.6945 |
| | 2 | 13.6815 | 12.9152 | 12.3343 | 10.4206 | 10.2948 | 9.7328 |
| 0.1 | 0.6 | 56.8967 | 45.0923 | 35.0643 | 17.5196 | 13.9320 | 9.3561 |
| | 0.8 | 36.7592 | 29.8086 | 24.2330 | 13.8996 | 12.2549 | 9.4047 |
| | 1 | 26.7369 | 22.4260 | 19.0840 | 12.2606 | 11.3810 | 9.4458 |
| | 1.5 | 16.6455 | 15.0884 | 13.9085 | 10.7099 | 10.4404 | 9.5157 |
| | 2 | 13.2843 | 12.6022 | 12.0752 | 10.2054 | 10.0929 | 9.5560 |
| 0.2 | 0.6 | 41.7529 | 35.9058 | 30.4426 | 15.6184 | 12.9069 | 8.9206 |
| | 0.8 | 29.3166 | 25.3371 | 21.8089 | 12.7269 | 11.4787 | 8.9621 |
| | 1 | 22.5099 | 19.7988 | 17.5055 | 11.3931 | 10.7218 | 8.9997 |
| | 1.5 | 15.0147 | 13.9113 | 13.0250 | 10.1060 | 9.8971 | 9.0666 |
| | 2 | 12.3152 | 11.8061 | 11.3961 | 9.6782 | 9.5902 | 9.1061 |
| | | (m=1,n=2) | | | | | |
| 0.01 | 0.6 | 177.9298 | 146.8518 | 119.2771 | 72.4860 | 56.0200 | 23.4448 |
| | 0.8 | 103.6319 | 86.5136 | 71.4632 | 45.5817 | 36.8242 | 18.7406 |
| | 1 | 69.1986 | 58.5687 | 49.3045 | 32.9925 | 27.7042 | 16.0971 |
| | 1.5 | 35.3123 | 31.0513 | 27.4021 | 20.3073 | 18.2782 | 12.9648 |
| | 2 | 23.6327 | 21.5239 | 19.7322 | 15.7393 | 14.7549 | 11.6746 |
| 0.1 | 0.6 | 128.6491 | 115.0075 | 100.8535 | 61.9095 | 50.7671 | 21.7604 |
| | 0.8 | 83.7200 | 73.8005 | 64.0823 | 40.9868 | 34.2963 | 17.7395 |
| | 1 | 59.4801 | 52.3247 | 45.5845 | 30.4743 | 26.1910 | 15.4054 |
| | 1.5 | 32.5876 | 29.1998 | 26.1803 | 19.3498 | 17.6033 | 12.5711 |
| | 2 | 22.4260 | 20.6396 | 19.0840 | 15.1956 | 14.3272 | 11.3811 |
| 0.2 | 0.6 | 83.4706 | 79.8473 | 75.5553 | 46.7388 | 41.9067 | 19.0838 |
| | 0.8 | 57.2834 | 55.5966 | 51.5392 | 33.3139 | 29.6359 | 16.0212 |
| | 1 | 45.0569 | 41.7813 | 38.3847 | 25.8975 | 23.2429 | 14.1341 |
| | 1.5 | 27.3400 | 25.3235 | 23.4005 | 17.3888 | 16.1546 | 11.7505 |
| | 2 | 19.7988 | 18.6005 | 17.5055 | 13.9934 | 13.3463 | 10.7218 |
| | | (m=2,n=1) | | | | | |
| 0.01 | 0.6 | 89.6437 | 76.4781 | 66.8141 | 46.8063 | 44.4696 | 38.6064 |
| | 0.8 | 65.1670 | 59.2380 | 54.8458 | 43.2079 | 42.2172 | 38.7873 |
| | 1 | 54.6743 | 51.6210 | 49.3045 | 41.6472 | 41.1469 | 38.9043 |
| | 1.5 | 45.3888 | 44.5274 | 43.8305 | 40.2293 | 40.0898 | 39.0679 |
| | 2 | 42.5528 | 42.2071 | 41.9144 | 39.7874 | 39.7326 | 39.1526 |
| 0.1 | 0.6 | 73.8831 | 66.4133 | 60.2869 | 42.7130 | 41.1136 | 36.1233 |
| | 0.8 | 57.0503 | 53.3087 | 50.3100 | 39.9417 | 39.2519 | 36.3047 |
| | 1 | 49.2606 | 47.2245 | 45.5845 | 38.7128 | 38.3610 | 36.4246 |
| | 1.5 | 41.9700 | 41.3592 | 40.8467 | 37.5793 | 37.4806 | 36.5941 |
| | 2 | 39.6410 | 39.3897 | 39.1713 | 37.2242 | 37.1858 | 36.6824 |
| 0.2 | 0.6 | 54.3699 | 51.5078 | 48.9119 | 35.7105 | 34.8930 | 31.1882 |
| | 0.8 | 44.8507 | 43.2486 | 41.8341 | 33.8972 | 33.5429 | 31.3334 |
| | 1 | 40.1384 | 39.2032 | 38.3847 | 33.0747 | 32.8922 | 31.4338 |
| | 1.5 | 35.4346 | 35.1279 | 34.8558 | 32.3003 | 32.2485 | 31.5782 |
| | 2 | 33.8397 | 33.7085 | 33.5896 | 32.0545 | 32.0309 | 31.6538 |
| | | (m=2,n=2) | | | | | |
| 0.01 | 0.6 | 201.2858 | 173.1369 | 148.7444 | 103.9515 | 90.1967 | 57.7020 |
| | 0.8 | 127.9210 | 113.4015 | 100.9808 | 76.0508 | 69.1671 | 50.4009 |
| | 1 | 94.3686 | 85.9792 | 78.8455 | 62.8595 | 58.9430 | 46.6393 |
| | 1.5 | 62.2293 | 59.3935 | 56.9663 | 49.6579 | 48.3432 | 42.6228 |
| | 2 | 51.6210 | 50.3833 | 49.3045 | 45.0551 | 44.4758 | 41.1469 |
| 0.1 | 0.6 | 128.6491 | 132.6772 | 121.7703 | 85.6022 | 77.8972 | 51.6274 |
| | 0.8 | 101.3485 | 94.0851 | 87.2358 | 65.9054 | 61.6058 | 45.9146 |
| | 1 | 79.1951 | 74.4019 | 70.0219 | 55.9735 | 53.3852 | 42.8870 |
| | 1.5 | 55.6450 | 53.7714 | 52.1001 | 45.5302 | 44.6033 | 39.5886 |
| | 2 | 47.2245 | 46.3599 | 45.5845 | 41.7409 | 41.3217 | 38.3610 |
| 0.2 | 0.6 | 93.8549 | 91.2142 | 88.3406 | 63.0974 | 60.3845 | 42.1865 |
| | 0.8 | 71.6052 | 69.2883 | 66.9264 | 51.3549 | 49.5471 | 38.2999 |
| | 1 | 59.1227 | 57.3380 | 55.5860 | 45.0445 | 43.8579 | 36.1646 |
| | 1.5 | 44.7624 | 43.9183 | 43.1236 | 38.0442 | 37.5757 | 33.7882 |
| | 2 | 39.2032 | 38.7801 | 38.3847 | 35.3839 | 35.1634 | 32.8922 |